\documentclass[12pt,preprint]{aastex}

\begin{document}

\title{POWER SPECTRUM AND INTERMITTENCY OF Ly${\alpha}$
  TRANSMITTED FLUX OF QSO HE2347-4342}

\author{Priya Jamkhedkar\altaffilmark{1},
Long-Long Feng\altaffilmark{2,3}, Wei Zheng\altaffilmark{4} and
Li-Zhi Fang\altaffilmark{1}}

\altaffiltext{1}{Department of Physics, University of Arizona,
Tucson, AZ 85721}
\altaffiltext{2}{National Astronomical Observatories, Chinese
Academy of Science, Chao-Yang District, Beijing, 100012, P.R.
China}
\altaffiltext{4}{Department of Physics and Astronomy, Johns
   Hopkins University, MD 21218-2686}

\begin{abstract}

We have studied the power spectrum and the intermittent behavior
of the fluctuations in the transmitted flux of HE2347-4342 ${\rm
Ly}{\alpha}$ absorption in order to investigate if there is any
discrepancy between the LCDM model with parameters given by the
WMAP and observations on small scales. If the non-Gaussianity of
cosmic mass field is assumed to come only from halos with an
universal mass profile of the LCDM model, the non-Gaussian
behavior of mass field would be effectively measured by its
intermittency, because intermittency is a basic statistical
feature of the cuspy structures. We have shown that the Ly$\alpha$
transmitted flux field of HE2347-4342 is significantly
intermittent on small scales. With the hydrodynamic simulation, we
demonstrate that the LCDM model is successful in explaining the
power spectrum and intermittency of ${\rm Ly}{\alpha}$ transmitted
flux. Using statistics ranging from the second to eighth order, we
find no discrepancy between the LCDM model and the observed
transmitted flux field, and no evidence to support the necessity
of reducing the power of density perturbations relative to the
standard LCDM model up to comoving scales as small as about
$0.08~{\rm h}^{-1}~{\rm Mpc}$. Moreover, our simulation samples
show that the intermittent exponent of the Ly$\alpha$ transmitted
flux field is probably scale-dependent. This result is different
from the prediction of universal mass profile with a constant
index of the central cusp. The scale-dependence of the
intermittent exponent indicates that the distribution of baryonic
gas is decoupled from the underlying dark matter.

\end{abstract}


\section{INTRODUCTION}

The standard LCDM cosmogony is gaining more and more support from
observations of cosmic structures, including the temperature fluctuations
of cosmic background radiation, the clustering of galaxies and clusters,
and the transmission flux fields of QSOs' Ly$\alpha$ absorption
spectrum. In the linear regime, the power spectrum of mass density
perturbations predicted by the
LCDM model  is found to be consistent with observations
on scales from few thousand to about 1 h$^{-1}$ Mpc (Pope et al. 2004).
In the non-linear regime, N-body simulations of the standard LCDM
model reveal that the mass density profile of dark matter halos is
probably universal, and the cosmic mass field can be modeled as a
superposition of the universal halos on various mass scales (e.g.
Cooray \& Sheth, 2002). This universal halo scenario has been
successful in describing the second and higher order correlations
of the evolved mass fields. It has also  been extensively
applied to model the formation and evolution of galaxies using the mass
function, the universal density profile, the two-point correlation of
the host halos, and the bias model of the relevant objects.

However, whether the LCDM model explains observations on sub-Mpc
scales is still unclear. The universal density profile of dark matter
halos given
by LCDM N-body simulation is cuspy or singular, i.e., with a
central density distribution given by $\rho(r)\propto r^{-\alpha}$,
where $\alpha=1$ (Hernquist, 1990; Navarro et al. 1996; Jing
2000), or $3/2$ (Moore et al. 1999). However, the mass density
profile given by the rotation curves of dwarf and low surface
brightness galaxies generally show a soft core in their centers.
The best fitted profiles are not in general as dense in the
predicted cuspy center (Flores \& Primack 1994; Swaters et al
2003; McGaugh et al. 2003; Zentner \& Bullock 2003; Simon et al.
2004). Moreover, the observed substructures within halos are lower
than predicted. These discrepancies have been used to prove that
the power of density perturbations on small scales is less than
the LCDM model prediction. This result has already motivated
attempts to modify the LCDM model on small scales, such as the warm
dark matter model, annihilating cold dark matter model (Kaplinghat et al
2000), and self-interacting dark matter model (Spergel \& Steinhardt
2000).

On the other hand, the
observed gravitational lensing of galaxy clusters, which yield
constraints on small scale behavior of structure clustering, are
found in good agreement with the LCDM model prediction (Metcalf
2004, Natarajan \& Springel 2004). No reduction of small scale
power is needed. Also, the $N$-body simulations show that no more
than 70\% of halos can be fitted by the standard universal
spherical mass profile. There is considerable amount of variation
in the density profile even among the halos that can be fitted by
the universal mass profile (Jing 2000; Bullock et al. 2001;
Ricotti 2003). The variation of the concentration parameter can
be as large as a factor of two. Therefore, one needs more tests on
the possible discrepancy between observations and prediction on
small scales, especially using high quality samples.

In this paper, we study the small scale behavior of the cosmic
field using the ${\rm Ly\alpha}$ transmitted flux of the QSO
HE2347-4342, and model samples given by the hydrodynamic
simulation. Along with the power spectrum, we also focus on the
intermittent behavior of the field of QSO's ${\rm Ly}{\alpha}$
transmitted flux on small scales. Roughly speaking, the
intermittency of random field is characterized by strong
enhancements (cuspy structures) scattered in a space with a low
density background. If the cosmic mass field is given by the
superposition of universal halos on various scales, all the
non-Gaussian features should be from the halos. Thus, the
intermittent features of QSOs' Ly$\alpha$ transmitted flux on
small scales would be effective to detect the cuspy behavior of
cosmic mass field. This approach is not new. The intermittency of
${\rm Ly}\alpha$ transmitted flux has been studied using the
absorption spectra of QSOs (Jamkhedkar et al. 2000; Jamkhedkar et
al. 2003). The results have also been used to compare with
simulations of standard LCDM model and warm dark matter model
(Pando et al 2002; Feng et al. 2003). However, there are two
reasons we want to revisit this topic. First, the data of
HE2347-4342 ${\rm Ly}{\alpha}$ transmitted flux have higher
resolution and S/N ratio than the Keck data used in previous works
(Pando et al 2002; Jamkhedkar et al. 2003). Second, the newly
developed hybrid cosmological hydrodynamic based on Weighted
Essentially non-Oscillatory scheme (WENO) is especially effective
in capturing singular and complex structures with a higher order
spatial accuracy (Feng et al 2004). ${\rm Ly}{\alpha}$ transmitted
flux has been extensively studied to calculate the power spectrum
of mass perturbation on small scales (Croft et al. 2002; Viel et
al 2004; McDonald et al. 2004).

The paper is organized as follows. In \S 2, we address the
statistics of intermittency. Section 3 describes the data of
HE2347-4342 and \S 4 presents the samples given by the WIGEON
simulation. Sections 5 and \S 6 show the analysis and
comparison of the power spectra and intermittent properties of
observed data and simulation samples, respectively. The discussion and
conclusion is presented in \S 7.

\section{INTERMITTENT MASS FIELD}

\subsection{Cuspy halos and intermittency}

Cuspiness of the mass density distribution can effectively be
described up by density difference $\Delta \rho_{r}({\bf x})
\equiv |\rho({\bf x +r})-\rho({\bf x})|$, where $r=|\bf r|$. For a
field given by superposition of cuspy halos,   the field is
regular  at most locations {\bf x}, i.e. $\Delta \rho_{ r}({\bf
x})\rightarrow 0$, when $r\rightarrow 0$. On the other hand, cusps
yield singular behavior,i.e.,  $\Delta \rho_{r}({\bf x})\rightarrow
\infty $, when $r\rightarrow 0$. That is, the mass field consists
of high spikes randomly and widely scattered in space, with a low
field value between the spikes. Such spiky field is generally
intermittent.

For a statistically homogeneous and isotropic random field, the
probability distribution function (PDF) of density difference
$\Delta \rho_{r}({\bf x})$ for a given scale $r$ has to be
independent of ${\bf x}$.  The existence of singular density
profiles means that events with large density difference
$\Delta \rho_{r}$ on small $r$ are more frequent than compared
with a Gaussian field. Therefore, the PDF of $\Delta \rho_{r}$ has
to be long tailed, given by the events with extremely large
density difference $\Delta \rho_{r}$. The long tail would be more
prominent for the PDF of smaller scales $r$. Thus, the long tail
of $\Delta \rho_{r}$ PDF of the mass field is an alternative tool to
probe the cuspiness of the  field.

As higher order moments are sensitive to the tail of the PDF,
an effective measurement of the PDF long tail is given by the
so-called structure function defined as
\begin{equation}
\label{S2ndef}
S^{2n}_r \equiv  \langle|\Delta \rho_r({\bf x})|^{2n} \rangle.
\end{equation}
where $\langle...  \rangle$ is the average over the ensemble of
fields. As above mentioned, if the field is statistically
homogeneous, $S^{2n}_r$ is independent of ${\bf x}$ and depends
only on $r$. When $n=1$, we have $S^2_r =\langle |\Delta
\rho_r({\bf x)}|^{2} \rangle$, which is the mean of the square of
the density fluctuations at $r$, and therefore, $S^2_r$ actually
is the power spectrum of mass density fluctuations of the field.

Intermittency of a random field is defined by the divergence of the
following ratio
\begin{equation}
\label{S2nratioandzeta}
\frac{S^{2n}_r}{[S^{2}_r]^n} \propto
   \left(\frac{r}{L} \right )^{-\zeta},
\end{equation}
where $L$ is the size of the sample, and $\zeta$ is called
intermittent exponent. Generally, $\zeta$ is $n$- and
$r$-dependent. The ratio in eq.~(\ref{S2nratioandzeta}) is the
${2n}^{\rm th}$ moment, $S^{2n}_r$, normalized by the power
$S_r^2$. As $S^{2}_r$ measures the ``width" (variance) of the PDF
of $\Delta \rho_r(x)$, and $S^{2n}_r$ is sensitive to the tail of
the PDF, the ratio in eq.~(\ref{S2nratioandzeta}) measures the
fraction of events in the long tail on the scale $r$. If the exponent
$\zeta$ is zero or negative, the field is regular, i.e. smooth on
smaller scales. If $\zeta$ is positive, the ratio diverges as
$r\rightarrow  0$, and the field is rough on small scales.  In
this case, the field is called to be intermittent (G\"artner \&
Molchanov, 1990; Zel'dovich, Ruzmaikin, \& Sokoloff, 1990). Since
$\Delta \rho_r({\bf x}) \rightarrow 0$ for regular field, the
$r\rightarrow  0$ asymptotic behavior of $S^{2n}_r/[S^{2}_r]^n$ is
dominated by cuspy structures. Intermittent exponent $\zeta$
measures cuspiness of the field. If the cuspy behavior is given by
$\rho(r)\propto r^{-\alpha}$ with a constant $\alpha$, the
exponent $\zeta$ should also be $r$-independent.

For a Gaussian field $\rho({\bf x})$, the PDF of the
density difference $\Delta \rho_r$ is also Gaussian. We have then
\begin{equation}
\label{S2nratiogaussian}
\frac{S^{2n}_r}{[S^{2}_r]^n} = (2n-1)!!.
\end{equation}
This ratio is independent of scale $r$, and therefore, the
intermittent exponent $\zeta=0$.

\subsection{Intermittent statistics with DWT variables}

The basic statistical variable in eqs.~(\ref{S2nratioandzeta}),
~(\ref{S2nratiogaussian}) is the density
difference $\rho({\bf x+r})-\rho({\bf x})$, which contains the
information of position ${\bf x}$ and spatial scale $r$.
Therefore, it is convenient to use the statistical variables given
by the discrete wavelet transform (DWT) decomposition of density
field. For an 1-D field, the quantity $\rho(x+r)-\rho(x)$ in terms of
DWT variables is
given by the wavelet function coefficient (WFC) defined as
\begin{equation}
\label{wfcdef}
\tilde{\epsilon}_{j,l} =\int \psi_{j,l}(x)\rho(x)dx,
\end{equation}
where $\psi_{j,l}(x)$ is the discrete wavelet basis function, $j$ denotes
the scale $L/2^j$, and $l$ the spatial range $lL/2^j$ to $(l+1)L/2^j$
(Daubechies, 1992; Fang \& Thews 1998). The WFC,
$\tilde{\epsilon}_{j,l}$, is the density fluctuation (or
difference) on scale $L/2^j$ at position $l$.

The structure functions in eq.~(\ref{S2ndef}) can then be re-written
via the WFCs as (Farge at al. 1996)
\begin{equation}
\label{S2ninwfcs}
S^n_j = \langle|\tilde{\epsilon}_{j,l}|^n \rangle.
\end{equation}
The fair sample hypothesis allows one to calculate
$S_j^n$ by using spatial averages, i.e., the average over $l$ so
that
\begin{equation}
\label{S2nwfcensembleaverage} S^n_j=
\frac{1}{2^j}\sum_{l=0}^{2^j-1}|\tilde{\epsilon}_{j,l}|^n.
\end{equation}
For $n=2$, we have
\begin{equation}
\label{S2nn2waveletpowerspectrum} S^2_j
=\frac{1}{2^j}\sum_{l=0}^{2^j-1}|\tilde{\epsilon}_{j,l}|^2.
\end{equation}
which is actually the power spectrum in the DWT modes $P_j\equiv
S^2_j$ (Pando \& Fang 1998; Fang \& Feng 2000). For a Gaussian
field, the Fourier power spectrum $P(n)$ is related to its DWT
power spectrum $P_j$ by
\begin{equation}
\label{DWTpowerspectrumvsFFT} P(n) =\frac{1}{L}
\sum_{j=0}^{\infty}P_j
 \left |\hat{\psi} \left (\frac {n}{2^j}\right) \right |^2,
\end{equation}
or
\begin{equation}
P_j = \frac{1}{2^j} \sum_{n = - \infty}^{\infty}
|\hat{\psi}(n/2^j)|^2 P(n),
\end{equation}
where $\hat{\psi}$ is the Fourier transform of the wavelet
function.  This implies that the DWT power spectrum $P_j$ is the
banded Fourier power of the flux fluctuations, and the band $j$
corresponds to the wavenumber around $k = 2 \pi n/L \simeq 2\pi
2^j/L$.

Since $r=L/2^j$, the intermittent exponent $\zeta$ defined by eq.
(\ref{S2nratioandzeta}) can be calculated by
\begin{equation}
\label{S2nratioandzetawavelets} \frac{S^{2n}_j}{[S^2_j]^n} \propto
2^{j\zeta}.
\end{equation}
Generally, $\zeta$ depends on $n$ and $j$. The statistics of the
power spectrum and intermittency are based entirely on the DWT
variables. It would be easy to make a uniform comparison between
model predictions and observations with statistics on second and
higher orders.

\section{DATA OF HE2347-4342}

The data used in our analysis is the transmitted flux of the ${\rm
Ly}\alpha$ absorption spectrum of QSO HE2347-4342 ($z=2.885$,
$V=16.1$).  The optical echelle spectra were obtained at the ESO
VLT UVES on 2001 November 23-24. The details on HE2347-4342
optical spectra have been described in Zheng et al 2004. The VLT
data cover the wavelength range between 3600 and 4800~\AA, which
corresponds to the entire ${\rm Ly}\alpha$ wavelength range
studied with FUSE from $z = 2.0$ to $2.9$. Using IRAF tasks
designed for echelle data, a normalized spectrum was obtained. The
spectrum has $24000$ points with resolution $\delta \lambda \simeq
0.05$ \AA. The data are given in the form of pixels with
wavelength $\lambda$, flux $F$ and noise $\sigma$. In terms of the
local velocity the resolution is $d v \simeq 3.5$ km~s$^{-1}$. The
S/N ratio of the spectrum is about $110$ per 0.1~\AA\ bin at
4700~\AA, and about $46$ at 3850~\AA. This is respectively $\sim$
$2.5$ and $10$ times of the data of Keck echelle spectrum used in
Feng et al 2003.

For our purpose, the useful wavelength region is from ${\rm
Ly}\beta$ absorption to the ${\rm Ly}\alpha$ emission, excluding a
region close to the quasar to avoid proximity effects. Below
3984~\AA~${\rm Ly}\beta$ absorption starts to appear. Therefore,
we take the range from $3986.01 - 4395.600$~\AA, corresponding to
redshift from $2.278$ to $2.615$. In this wavelength range,
the mean transmission $\langle e^{-\tau} \rangle$ is $0.796$.
This redshift range contains about 2$^{13}$ pixels. The size of
a cell on the DWT scale $j$
corresponds to $N=2^{13-j}$ pixels. The distance between $N$
pixels in the units of the local velocity scale is given by
$\delta v=2c[1-\exp(-N d v/2c)]$~km~s$^{-1}$,
corresponding to comoving scale $D = (\delta v
/H_0)[\Omega_m(1+z_m)^3+\Omega_{\Lambda}]^{-1/2}$.

Metal lines are a major cause of contamination in the spectra. But
the doppler width of metal lines are generally narrow with $\delta
v \leq~ 20~{\rm km~s^{-1}}$. In this paper, we restrict our
analysis only to scales $\delta v \geq 30~{\rm km~s^{-1}}$ where
contamination due to metal lines is low (Hu et al 1995;
Boskenberg et al 2003; Kim, et al. 2004).

\section{HYDRODYNAMIC SIMULATION}

The simulation uses the newly developed hybrid cosmological
hydrodynamic codes based on the Weighted Essentially
Non-Oscillatory (WENO) scheme (Harten et al. 1987; Liu et al.
1994; Jiang \& Shu 1996; Shu 1998; Fedkiw et al. 2003; Shu 2003).
We will name this code as WIGEON, {\bf W}eno for {\bf
I}ntergalactic medium and {\bf G}alaxy {\bf E}volution and
formati{\bf ON}. For details of the numerical method and tests, we
refer to Feng et al.(2004). The simulation sample we analyze here
is actually the same as those used in our previous papers on the
statistical study of temperature, entropy, baryonic fraction and
velocity fields of intergalactic medium (He et al, 2004; He et al.
2005; Kim et al. 2005). It was performed in a cubic box of side
length 12 h$^{-1}$ Mpc with a 192$^3$ grid and an equal number of
dark matter particles. The cosmogony is the standard LCDM model
specified by the density parameter $\Omega_m=0.3$, the baryon
density $\Omega_b=0.047$, the cosmological constant
$\Omega_{\Lambda}=0.7$, the Hubble constant $h=0.7$, the shape
factor
$\Gamma=\Omega_{m}h\exp[-\Omega_b(1+\sqrt{2h}/\Omega_m)]=0.166$,
and the rms mass fluctuations in spheres of $8h^{-1}$Mpc,
$\sigma_8=0.9$. The ratio of specific heats is $\gamma=5/3$. Since
the shock heating of cosmic gas is significant (He et al 2004),
the resolution of the simulation should be less than the thickness
of the shock, which is of the order of the dissipation length,
i.e., the Jeans diffusion $\sim 0.1 - 0.3$ h$^{-1}$ Mpc for
redshifts $z<4$ (Bi et al. 2003).  The size of the grid is
$12/192=33/2^8=0.063$ h$^{-1}$ Mpc. Therefore, the resolution of
our simulation is sufficient to capture shocks.

Atomic processes including ionization, radiative cooling and
heating are modelled as in Cen (1996) in a primeval plasma of
hydrogen and helium of composition ($X=0.76$, $Y=0.24$). The
uniform UV-background of ionizing photons is assumed to have a
power-law spectrum of the form $J(\nu) =J_{21}\times
10^{-21}(\nu/\nu_{HI})^{-\alpha}$ergs$^{-1}$cm$^{-2}$sr$^{-1}
$Hz$^{-1}$, with $\alpha=1$, where the photo ionizing flux is
normalized by parameter $J_{21}$ at the Lyman limit frequency
$\nu_{HI}$, and is suddenly switched on at $z\sim 6$ to heat the
gas and re-ionize the universe.

One-dimensional fields are extracted along randomly selecting
lines of sight in the simulation box. The density, temperature and
velocity of the neutral gas fraction on grids are Gaussian
smoothed using FFT techniques which form the fundamental data set.
The one-dimensional grid containing the physical quantities is
further interpolated by a cubic spline. Using this one-dimensional
grid, the optical depth $\tau$ is then obtained by integrating in
real space and we include the effect of the peculiar velocity and
convolve with Voigt thermal broadening. To have a fair comparison
with observed spectra, $\tau$ was Gaussian smoothed to match with
the spectral resolutions of observation. The transmitted flux
$F=\exp(-\tau)$ is normalized such that the mean flux decrement in
the spectra match with observations.

Each mock spectrum is sampled on a $2^{10}$ grid with the same
spectral resolution as the observation. As the corresponding
comoving scale for $2^{10}$ pixels is larger than the simulation
box size, we replicate the sample periodically. To achieve the
greatest statistical independence, we randomly change the
direction of line of sight while crossing the boundary of the
simulation box. By the way, 1000 mock spectra are generated.

\section{POWER SPECTRUM OF THE TRANSMITTED FLUX OF HE2347-4342}

\subsection{Treatment of unwanted modes}

In order to calculate the DWT power spectrum
[eq.~(\ref{S2nn2waveletpowerspectrum})] of the transmitted flux of
HE2347-4342, we should properly treat unwanted data, including the
pixels without data, contamination of metal lines etc. Although
the $S/N$ is high on an average, it is as low as about 1 for some
pixels, such as pixels with negative flux. We must reduce the
uncertainty given by low S/N pixels. In the DWT analysis, the
conventional technique of reducing these uncertainties is given by
the algorithm of DWT denoising by thresholding (Donoho 1995) or
conditional-counting, (Jamkhedkar et al. 2003) as follows
\begin{enumerate}
\item Calculate the SFCs of both transmission $F(x)$ and
noise $\sigma(x)$, i.e.
\begin{equation}
\label{DWToffluxandnoise}
\epsilon^F_{jl}=\int F(x)\phi_{jl}(x)dx, \hspace{3mm}
\epsilon^N_{jl}=\int \sigma(x)\phi_{jl}(x)dx.
\end{equation}
\item Identify an unwanted mode $(j,l)$ using the threshold condition
\begin{equation}
\label{conditionalcounting}
\left| \frac{\epsilon^F_{jl}} {\epsilon^N_{jl}} \right| <  f
\end{equation}
where $f$ is a constant. This condition flags all modes with S/N
less than $f$. We can also flag modes dominated by metal lines.
\item Since all the statistical quantities in the DWT
representation are based on an average over the modes $(j,l)$, we
will skip all the flagged modes while computing these averages,
i.e. the average is over the un-flagged modes $N(f)$ only.
\end{enumerate}
With this method, no rejoining and smoothing of the data are
needed. The condition in eq.~(\ref{conditionalcounting}) is
applied on each scale $j$, and therefore the unwanted modes are
flagged on a scale-by-scale basis.  Generally, for scale $j$,
$N(f)\leq 2^j$. If the size of an unwanted data segment is $R$,
condition in eq.~(\ref{conditionalcounting}) only flags modes $(j,l)$
on scales less than or comparable to $R$. We also flag two modes
around each unwanted mode to reduce any boundary effects of the
chunks. With the conditional-counting method, we can still
calculate the power spectrum by the estimators of
eq.~(\ref{S2nn2waveletpowerspectrum}), but the average is not over
all modes $l$, but over the un-flagged modes only.

\subsection{Power spectrum of the DWT modes}

We calculate the power spectrum of the transmitted flux
fluctuations, $\Delta F=F(\lambda) - \langle F(\lambda)\rangle$,
of HE2347-4342. To consider the correction of the noise on
eq.~(\ref{S2nn2waveletpowerspectrum}), the
power spectrum of transmitted flux is given by (Pando \& Fang
1998b; Fang \& Feng 2000; Jamkhedkar, Bi, \& Fang, 2001)
\begin{equation}
\label{DWTpower}
P_j=\frac{1}{2^j}\sum_{l=0}^{N(f)}(\tilde{\epsilon}^F_{jl})^2
   - \frac{1}{2^j}\sum_{l=0}^{N(f)}(\tilde{\epsilon}^n_{jl})^2.
\end{equation}
The first term on the r.h.s. of eq.~(\ref{DWTpower}) is the same
as eq.~(\ref {S2nn2waveletpowerspectrum}), in which the wavelet
coefficients (WFC) $\tilde{\epsilon}^F_{jl}$ are given by
\begin{equation}
\label{fluxDWT}
 \tilde{\epsilon}^F_{jl} = \int F(x)\psi_{jl}(x)dx.
\end{equation}
The second term on the r.h.s. of eq.(\ref{DWTpower}) is due
to the noise field
$\sigma(\lambda)$ and is calculated by
\begin{equation}
\label{noiseDWT}
(\tilde{\epsilon}^n_{jl})^2 = \int\sigma^2(x)\psi_{jl}^2(x)dx.
\end{equation}

Figure 1 plots the results of the DWT power spectrum, in which the
parameter $f$ is taken to be 1, 2, 3 and 5. At the first glance, the
conditional-counting of eq.~(\ref{conditionalcounting}) would seem to
preferentially drop modes in the low transmission regions, and
power spectrum eq.~(\ref{DWTpower}) should be  $f$-dependent.
 However, Figure 1 shows that the power spectrum $P_j$ is
independent of $f$ on entire the scale range considered for $f=1$
to 5. This can be seen from eq.~(\ref{DWTpower}), which shows that
the contribution to the power $P_j$ given by mode $(j,l)$ is
$(\tilde{\epsilon}^F_{jl})^2-(\tilde{\epsilon}^n_{jl})^2$. The
noise substraction term $(\tilde{\epsilon}^n_{jl})^2$ guarantees
that the contribution of modes with small ratio $S/N$ to $P_j$ is
always small or negligible. For instance, the modes with negative
flux, i.e., the modes with flux having the same order of magnitude
as noise, the two terms $(\tilde{\epsilon}^F_{jl})^2$ and
$(\tilde{\epsilon}^n_{jl})^2$ statistically cancel each other.
Thus, all the dropped modes have a very small or negligible
contribution to $P_j$ regardless the parameter $f$. Denoising by
thresholding or conditional-counting is reliable.

Figure 2 compares the DWT power spectra measured in the mock
samples and the observed data. We take the same parameter $f=3$ for
both real data and mock samples. The error bars of $P_j$ are
the maximum and minimum range of $P_j$ from bootstrap re-sampling.
Since the PDF of $\tilde{\epsilon}^F_{jl}$ is highly non-Gaussian
(\S 5.1), a reasonable estimation of the errors for the average
over the ensemble of $\tilde{\epsilon}^F_{jl}$ is given by
bootstrap re-sampling (Jamkhedkar et al 2003). That is, for
observed sample, the bootstrap re-sampling is done on the set of
$N(f)$ data for each scale $j$, and for the simulation samples the
bootstrap re-sampling is on the set of $N_{sim}(f)$ data from the
1000 simulation samples. The error bar is calculated thus. For a data
set with $N$ points, $N$ realizations or data sets are created by
drawing points from the original set with replacement. The average is
calculated over the $N$ realizations. The power spectrum
is calculated on the  scales from $\delta v = 224$ to 28 km~s$^{-1}$,
corresponding to comoving scale $0.63 - 0.079$ h$^{-1}$ Mpc in the
LCDM model. The error bars for the real data are given by the
maximum and minimum of bootstrap re-sampling. Figure 2 shows that
simulation samples basically are in agreement with observations on
the scales considered. Particularly, no discrepancy has been found
even on the smallest scale $\delta v = 28~{\rm km~s^{-1}}$ or length
scale $D = 0.079$ h$^{-1}$ Mpc.

\section{INTERMITTENT BEHAVIOR}

\subsection{PDF of $\tilde{\epsilon}^F_{j l}$}

As first step to describe the intermittent behavior of the ${\rm
Ly}\alpha$ transmitted flux, we show in Figure 3 the PDFs of the
normalized WFCs of flux field $\tilde{\epsilon}^F_{j
l}/\langle(\tilde{\epsilon}^F_{j l})^2\rangle^{1/2}$. These
distributions are compared with a Gaussian distribution with zero
mean and unit standard deviation. On the scales $224$
km~s$^{-1}$ the PDFs of the normalized WFCs do not strongly
deviate from Gaussian fields. But on scales $<$ 224 km~s$^{-1}$,
the central peak and long-tail of the PDFs show that the field is
highly non-Gaussian. On the scale 56 km~s$^{-1}$, the long tail
extends to $\tilde{\epsilon}^F_{j l}/\langle(\tilde{\epsilon}^F_{j
l})^2\rangle^{1/2}\simeq 5$. That is, the power of some long tail
events $(\tilde{\epsilon}^F_{j l})^2$ can be larger than the mean
power $\langle(\tilde{\epsilon}^F_{j l})^2\rangle$ by a factor of
20-30. This is in agreement with the result based on Keck data
(Jamkhedkar et al 2003).

We see that for $f=1, 3$ and $5$, the PDFs given by $f=1, 3$ and 5
are essentially the same, especially, the central peaks of the
PDFs are insensitive to the parameter $f$. This indicates that the
statistical result does not depend on data at pixels with low
number of $S/N$. This point is important. For instance, a
saturated absorption region on scale $j_s$ may yield
$\tilde{\epsilon}^F_{j l}\rightarrow 0$ on smaller scales or
$j>j_s$, i.e. pixels with a low value of $S/N$. However, we cannot
draw information of clustering of cosmic matter from that region.
We also can not say whether this region underwent a strong
nonlinear evolution. Therefore, the $f$-independence of the PDFs
of Figure 3 provides an valuable measurement of the non-Gaussian
behavior, irrespective of whether the region is saturated or not.

\subsection{Structure functions}

We now calculate the structure functions
eqs.~(\ref{S2nratioandzeta}) or (\ref{S2nn2waveletpowerspectrum})
for the transmitted flux fluctuations of HE2347-4342. For real
data, the results are illustrated in Figure 4. In calculating the
high order moment $S^{2n}_j$, we did not subtract the noise term in
eq.~(\ref{fluxDWT}), because as noise is considered Gaussian, its
higher order moments are small (see \S 6.4 below). The error bars
are found by bootstrap re-sampling.  We see that $\log_2[
S^{2n}_j/(S^{2}_j)^n]$ with $n=2$ has errors even smaller than the
power spectrum of Figure 1. This is because the uncertainty in the
power spectrum is caused by rare and improbable long tail events
in the fluctuations, i.e. the tail of the PDF shown in Figure 3.
The large uncertainty of the PDF tail leads to the large
uncertainty of the power spectrum. On the other hand, the
structure function is the ratio between $S^{2n}_j$ and
$(S^{2}_j)^n$ and it reduces the effect of individual high spikes
(tail events). Therefore, the structure functions are an effective
and stable tool for high order statistics. Similar to the power
spectra of Figures 1 and 2, and the PDFs of Figure 3, the
structure functions of Figure 4 are basically independent of $f$.
Therefore, the scale- and $n$-dependencies of the structure
function exist regardless of the saturated absorption, and give a
measurement of the non-Gaussian clustering of the baryonic gas.

The value of $\log_2 [S^{2n}_j/(S^{2}_j)^2]$ generally is bigger
for smaller length scales. For a given $n$, the $j$-dependence of
$\log_2 [S^{2n}_j/(S^{2}_j)^n]$ can approximately be fitted by
eqs. (\ref{S2nratioandzeta}) and (\ref{S2nratioandzetawavelets})
with a positive exponent $\zeta$. Therefore, the field of the
transmitted flux fluctuations is highly intermittent. This shows
again that the cuspy feature of cosmic clustering can be seen in
high mass density areas, like massive halos, as well as in low
mass density areas, like the clouds of Ly$\alpha$ absorption
clouds (He et al. 2004; Pando et al. 2004; He et al. 2005; Kim et
al. 2005). Actually the success of the semi-analytical lognormal
model (Bi \& Davidsen, 1997) in explaining the ${\rm Ly}\alpha$
forest has already indicated that the mass field of the cosmic baryon
gas is probably intermittent, because a lognormal field is
intermittent.

The structure functions measured for the mock samples are shown in
Figure 5, in which the error bars are also given by bootstrap
re-sampling of the 1000 samples. It shows once again that the
statistical uncertainty of the structure function is reliable for
a high order statistical test. Figure 5 also shows that the
structure function is $f$-independent. Figure 6 gives a comparison
between the mock samples and the real data. We see that for all
scales from 224 to 28 km~s$^{-1}$, and all order $n$, the
structure functions of the mock samples are consistent with real
data within their error bars. The consistence is very good on the
smallest scale 28 km~s$^{-1}$.

\subsection{Intermittent exponent}

Following the definition of intermittent exponent $\zeta$
eq.~(\ref{S2nratioandzeta}) or eq.~(\ref{S2nratioandzetawavelets}),
we can calculate $\zeta$ in the scales range $j_1$ to $j_2$ by
\begin{equation}
\label{zetaDWT}
 \zeta_n = -\frac{1}{j_1-j_2}
\log_2\left[ \frac{S^{2n}_{j_1}(S^{2}_{j_2})^n}
{S^{2n}_{j_2}(S^{2}_{j_1})^n}  \right ].
 \end{equation}
The result is listed in Table 1. The error of $\zeta_n$ is
estimated by $\sigma_{\zeta_n} =
(1/|j_1-j_2|)\sqrt{{\sigma_1^2}+{\sigma_2^2}}$, where $\sigma_1$
and $\sigma_2$ are the errors of
$\log_2[{S^{2n}_{j_1}}/{(S^{2}_{j_1})^n}]$, and
$\log_2[{S^{2n}_{j_2}}/{(S^{2}_{j_2})^n}]$, respectively. The
scale $j$ represents a local velocity $\delta v = 2^{13-j} \times
3.5~$~km~s$^{-1}$. As expected, Table 1 shows that the values of
intermittent exponents of real data and mock samples are
consistent with each other on all orders and scales considered.
\begin{table}[htb]
\label{tab:zetavsj}
\caption{Intermittent exponent}
\bigskip
\begin{tabular}{clll}
\tableline
 {\rm scales} $j_1$ - $j_2$~(${\rm km~s^{-1}}$) & $2n$  &\multicolumn{2}{c}
{intermittent exponent   $\zeta_n$} \\ \hline
                     &  &
 {\rm real data} &
   {\rm mock sample}  \\
\tableline
 8 - 9 (112-56)  & 4 & $ 0.49\pm  0.21$ & $0.66\pm 0.03$ \\
 8 - 9  (112-56)  & 6 & $ 1.26\pm 0.48 $ & $1.54\pm 0.06$ \\
 8 - 9  (112-56) & 8 & $ 2.18\pm 0.75 $ & $2.50\pm 0.10$  \\
 9 - 10 (56-28) & 4 & $ 0.51\pm 0.32 $ & $0.51\pm 0.03$  \\
 9 - 10 (56-28) & 6 & $ 1.49\pm 0.76 $ & $1.08\pm 0.07$ \\
 9 - 10 (56-28) & 8 & $ 2.6\pm 1.2   $ & $1.67\pm 0.11$  \\
\tableline
\end{tabular}
\end{table}

In Table 2, we list the intermittent exponent of mock samples for
different scales ranges. It clearly shows that the intermittent
exponent is scale-dependent. We also tried to fit the
$j$-dependence of $\log_2 [S^{2n}_j/(S^{2}_j)^2]$ from $j=7$ to 10
by
\begin{equation}
\log_2 [S^{2n}_j/(S^{2}_j)^n]= A + j\zeta_n
\end{equation}
with assumption of $\zeta_n$ to be constant ($j$-independent). We
found that the goodness-of-fit, $Q$, (Press et al 1992) of the
data to eq.(18) with a constant $\zeta_n$ is always $\ll 0.1$.
That means that the assumption that $\zeta_n$ is independent of
scale does not hold. Therefore, $\zeta_n$ most likely is
scale-dependent. This result is inconsistent with the cuspy center
profiles $\rho(r) \propto r^{-\alpha}$ with a constant index
$\alpha$, which predicts a constant $\zeta_n$ on small scales.
Moreover, the $j$-dependence of $\zeta_n$ shown in Table 2 is not
monotonic. This makes it more difficult to fit with the standard
universal profiles. Therefore, the LCDM model seems to predict a
different scale-dependence of the intermittent exponent for the
Ly$\alpha$ absorption clouds in contrast to the cuspy behavior
of the universal profile of dark matter halos.
\begin{table}[htb]
\label{tab:zetaj}
\caption{Scale dependence of intermittent exponent}
\bigskip
\begin{tabular}{clll}
\tableline
 $2n$ & \multicolumn{3}{c}{$j_1-j_2$} \\
\tableline
  &\multicolumn{1}{c}{7-8} &\multicolumn{1}{c}{8-9} &
\multicolumn{1}{c}{9-10} \\
 \tableline
 4  & $0.563\pm 0.028$ & $0.66\pm 0.03$ & $0.51\pm 0.03$ \\
 6  & $1.183\pm  0.069$ & $1.54\pm 0.06$ & $1.08\pm 0.07$ \\
 8  & $1.77\pm  0.11$ &  $2.50\pm 0.10$ & $1.67\pm 0.11$ \\
\tableline
\end{tabular}
\end{table}

\subsection{$n$-dependence of structure function}

We now turn to the $n$-dependence of the structure functions.
Figures 7 and 8 are, respectively, $\log_2 [S^{2n}_j/(S^{2}_j)^n]$
vs. $n$ for the real and mock samples of HE2347-4342 on scales
$\delta v = 224-28~{\rm km~s^{-1}}$. For a Gaussian field, the
$n$-dependence of $\log_2(S^{2n}_j/(S^{2}_j)^n)$ is given by
eq.~(\ref{S2nratiogaussian}), i.e. $\log_2 (2n-1)!!$, which is also
plotted in Figures 7 and 8. The curves of $\log_2
[S^{2n}_j/(S^{2}_j)^n]$ vs. $n$ for both real and mock samples are
much higher than Gaussian field on scales $56~{\rm and}~28~{\rm
km~s^{-1}}$, but not much different from Gaussian field on scale
of 224 km~s$^{-1}$. Therefore, it is reasonable to ignore the
noise term in calculating the high order moment (\S 6.2).

More interesting is to fit the observed $n$-dependence for
$\log_2 [S^{2n}_j/(S^{2}_j)^n]$ with
\begin{equation}
\log_2 \frac{S^{2n}_j}{(S^{2}_j)^n} \propto n^{\alpha}(n-1).
\end{equation}
The motivation is to compare the fields with a lognormal field, for
which $\alpha=1$. Figure 7 shows that the best fit of $\alpha$ is in the
range $\alpha = 0.3$ - $0.4$ on scale $56~{\rm km~s^{-1}}$, and $0.3- 0.5$
on $28 ~{\rm km~s^{-1}}$. That is, the value of $\alpha$ seems to
approach to 1 when scale is small. The transmitted flux field is
closer to a lognormal field on small scales. This somewhat supports the
lognormal models of ${\rm Ly}\alpha$ forests. Figure 8 shows
that values of $\alpha$ given by mock samples always lie in the range
from $0.3-0.4$  and are consistent
with real data.

\section{DISCUSSIONS AND CONCLUSIONS}

We have showed that Ly$\alpha$ transmitted flux field of
HE2347-4342 is significantly intermittent, especially on small
scales. We found that the power spectrum is in good agreement with
the data of ${\rm Ly}{\alpha}$ transmitted flux of HE2347-4342.
There is no evidence of any discrepancy between the LCDM model
from observed intermittent features on scale as small as about
$\delta v = 28$ km~s$^{-1}$, and for statistical orders from 2 to
8. Accordingly, there is no need of reducing the power relative to
the standard LCDM model up to length scale $0.079$ h$^{-1}$ Mpc.

Comparing the current results with our previous studies on the
same topic, we found that the intermittency sensitively relies on
the quality of both the observed data and simulation sample. In
the first stage of our study, we studied the intermittency of
Ly${\alpha}$ transmitted flux with Keck data and model samples
produced by pseudo-hydro simulations (Pando et al 2002). Although
the simulation samples can fit the observed power spectrum, and
are also intermittent, the intermittent exponent does not fit the
real data. There is a discrepancy between the observed data and
simulation sample on small scales. Physically, that is probably
because the pseudo-hydro simulations assumed that (1) the baryon
distribution is proportional to that of dark matter
point-by-point, and (2) the gas temperature is related to the
density by a power law equation of state. However, it has been
shown that the relation between temperature and IGM density is
multi-phased. The relation between temperature and density can
approximately be described by a power-law equation. However, for a
given density, the temperature actually is not single-valued, but
varies from $10^4 - 10^7$ K (He et al 2004).

In the second stage, the model samples are produced by full hydro
simulations with two assumptions mentioned above (Feng et al
2003). The result was a great improvement with respect to the
first round. It shows that the intermittent behavior of the Keck
data and simulation is basically consistent with each other, but a
discrepancy can still be seen on scale $\delta v = 32$
km~s$^{-1}$. In the current study, the observed data of
HE2347-4342 probably is among the best quality for our purpose.
Its intermittency is in good agreement with the LCDM model on
small scales less than $\delta v = 30$ km~s$^{-1}$.

The star formation and their feedback on the cosmic gas evolution
are not considered in our simulation. Generally speaking, there
are two types of the feedbacks: (1) photoionization heating by the
UV emission of stars and AGNs, and (2) injection of hot gas and
energy by stars. The photoionization heating can be properly
considered, if the UV background is adjusted by fitting the
simulation with the observed mean flux decrement of QSO's ${\rm
Ly}{\alpha}$ absorption spectrum. The effect of injecting hot gas
and energy is localized in massive halos, and therefore, its
effect is weak when we  consider to avoid proximity effects.
Therefore, the major conclusions would not be significantly
affected even while considering the effect of star formation.

Intermittency is very effective to probe the details of the
singular features of a random field. Our simulation samples show
that the intermittent exponent of the Ly$\alpha$ transmitted flux
field probably is scale-dependent. This result is different from
the prediction of universal mass profile with cuspy center
$\rho(r) \propto r^{-\alpha}$. If the index $\alpha$ is constant,
the intermittent exponent should be scale-independent. Therefore,
the scale-dependence of the intermittent exponent indicates that
the distribution of baryon gas is decoupled from the underlying
dark matter (e.g. He et al 2004, Kim et al 2005). The data of
HE2347-4342 only is unable to test the prediction of
scale-dependence of the intermittent exponent. More high quality
QSO absorption spectra would be very valuable to test the $j$
dependence of $\zeta_n$ on small scales.

\acknowledgments

LLF acknowledges support from the National Science Foundation of
China (NSFC). This work is supported in part by the US NSF under
the grants AST-0507340.

\clearpage

\begin{figure}
\figurenum{1} \epsscale{1.0} \plotone{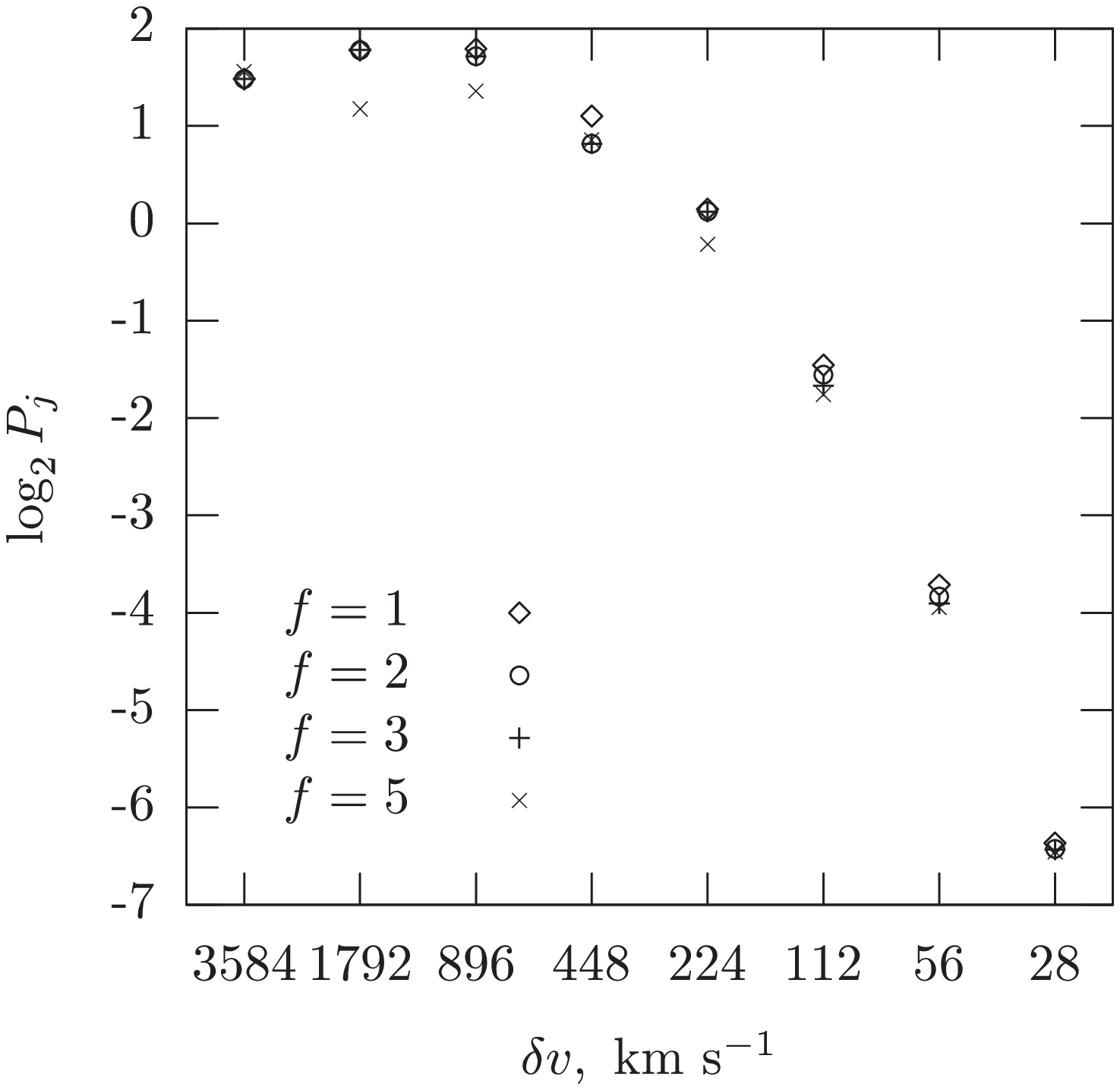} \caption{Power
spectrum of the HE2347-4342 transmitted flux for the conditional
counting parameter $f=1$ (diamond), 2 (circle), 3. (plus), and 5
(cross).}
\end{figure}

\begin{figure}
\figurenum{2} \epsscale{1.0} \plotone{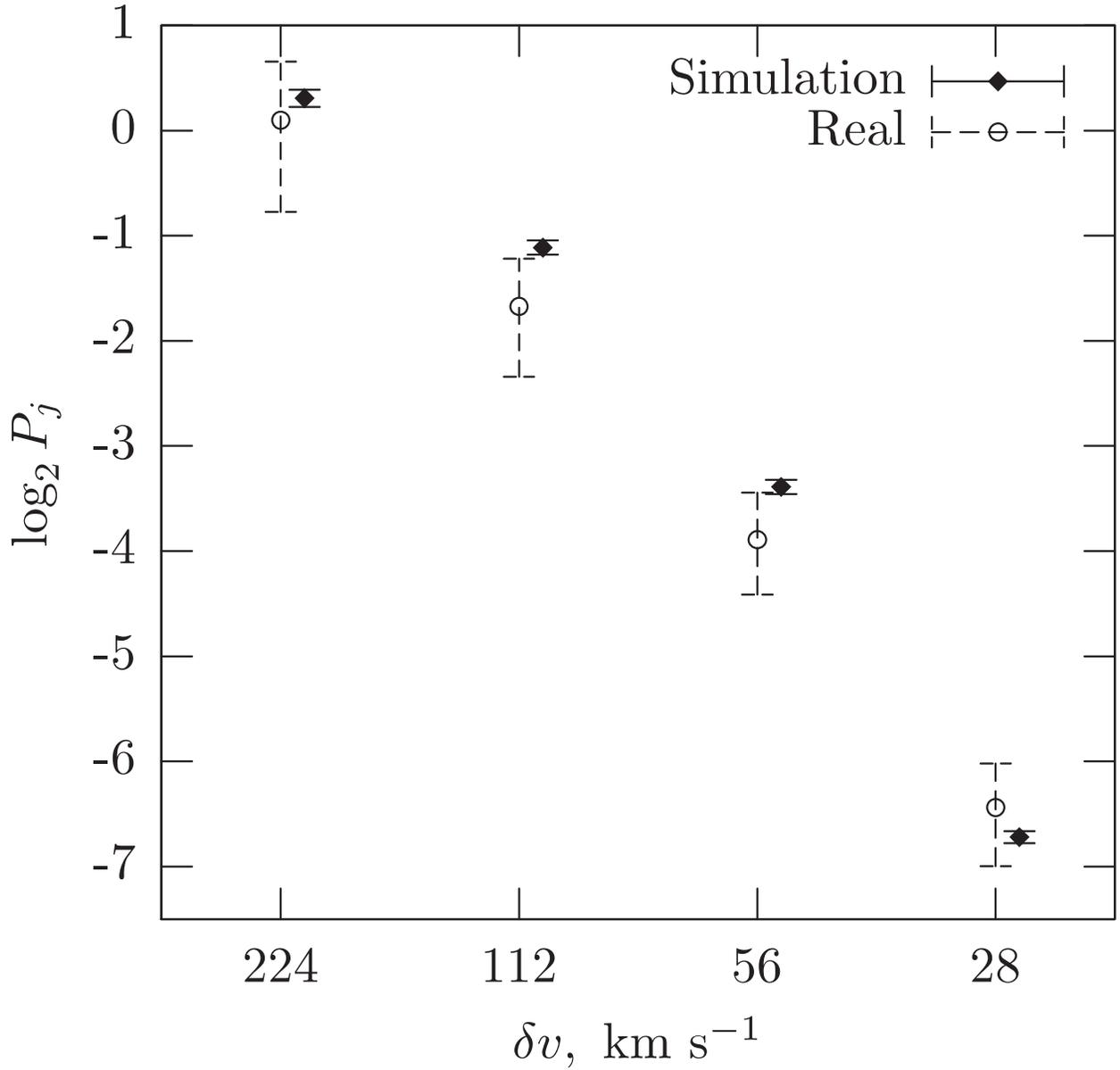}\vspace{-1.5cm}
\vspace{1.5cm} \caption{Power spectra of HE2347-4342 for 1.) real
data (circle) and 2.) mock samples (black diamond) by WIGEON
simulation with $f=3$. The error bars are the maximum and minimum
of bootstrap re-sampling. For clarity, the power spectrum of the
simulations is shifted slightly to the right.}
\end{figure}

\begin{figure}
\figurenum{3a} \epsscale{1.0} \plotone{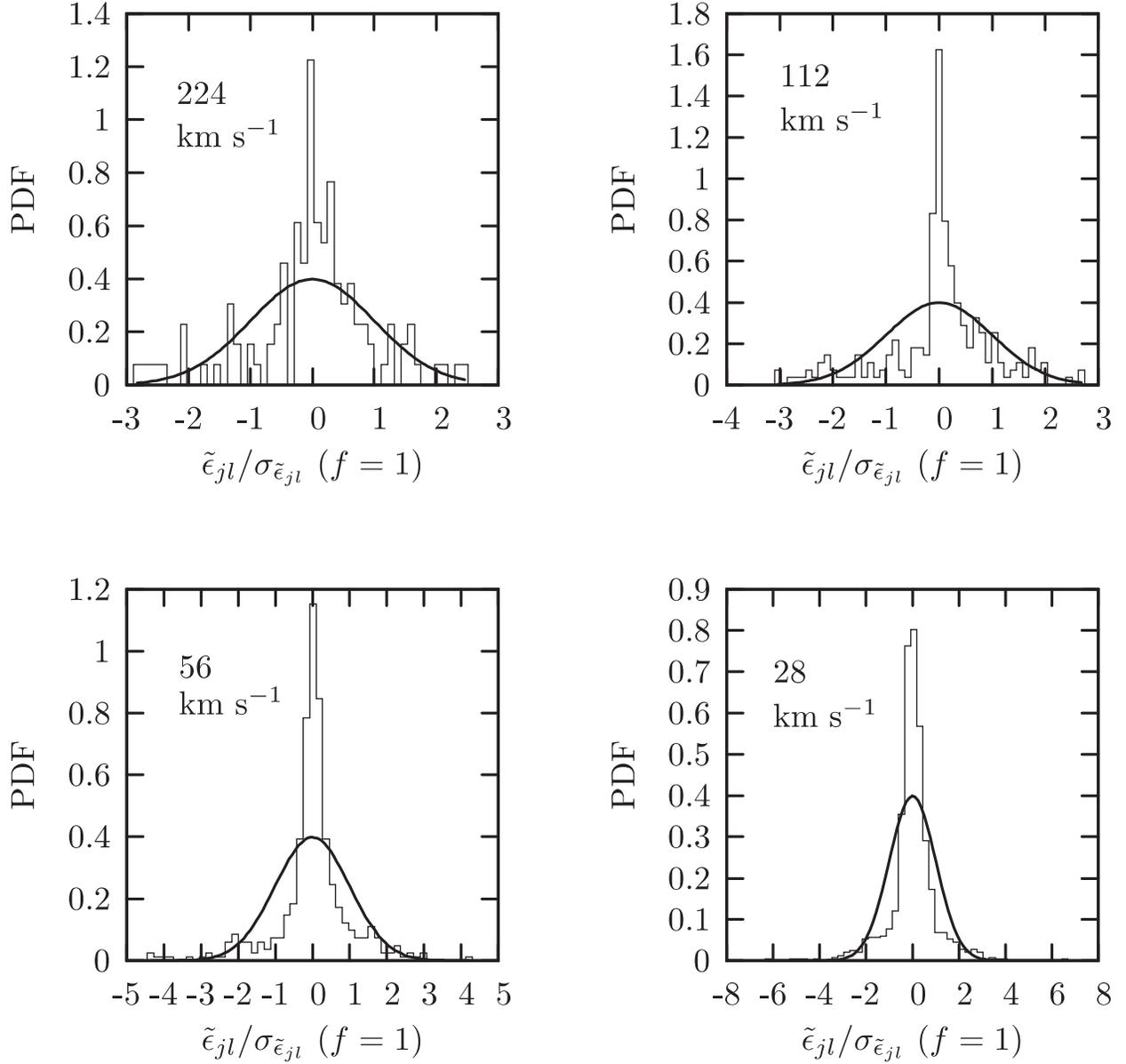} \caption{PDFs of
$\tilde{\epsilon}^F_{j l}/\langle(\tilde{\epsilon}^F_{j
l})^2\rangle^{1/2}$ for the scales $\delta v
=224,~112,~56,~28~{\rm km~s^{-1}}$. Figs. (a),(b) and (c)
represent the parameter $f=1,2,~{\rm and}~3$ respectively. A
Gaussian distribution with zero mean and unit standard deviation
is also showed in each panel. }
\end{figure}

\begin{figure}
\figurenum{3b} \epsscale{1.0} \plotone{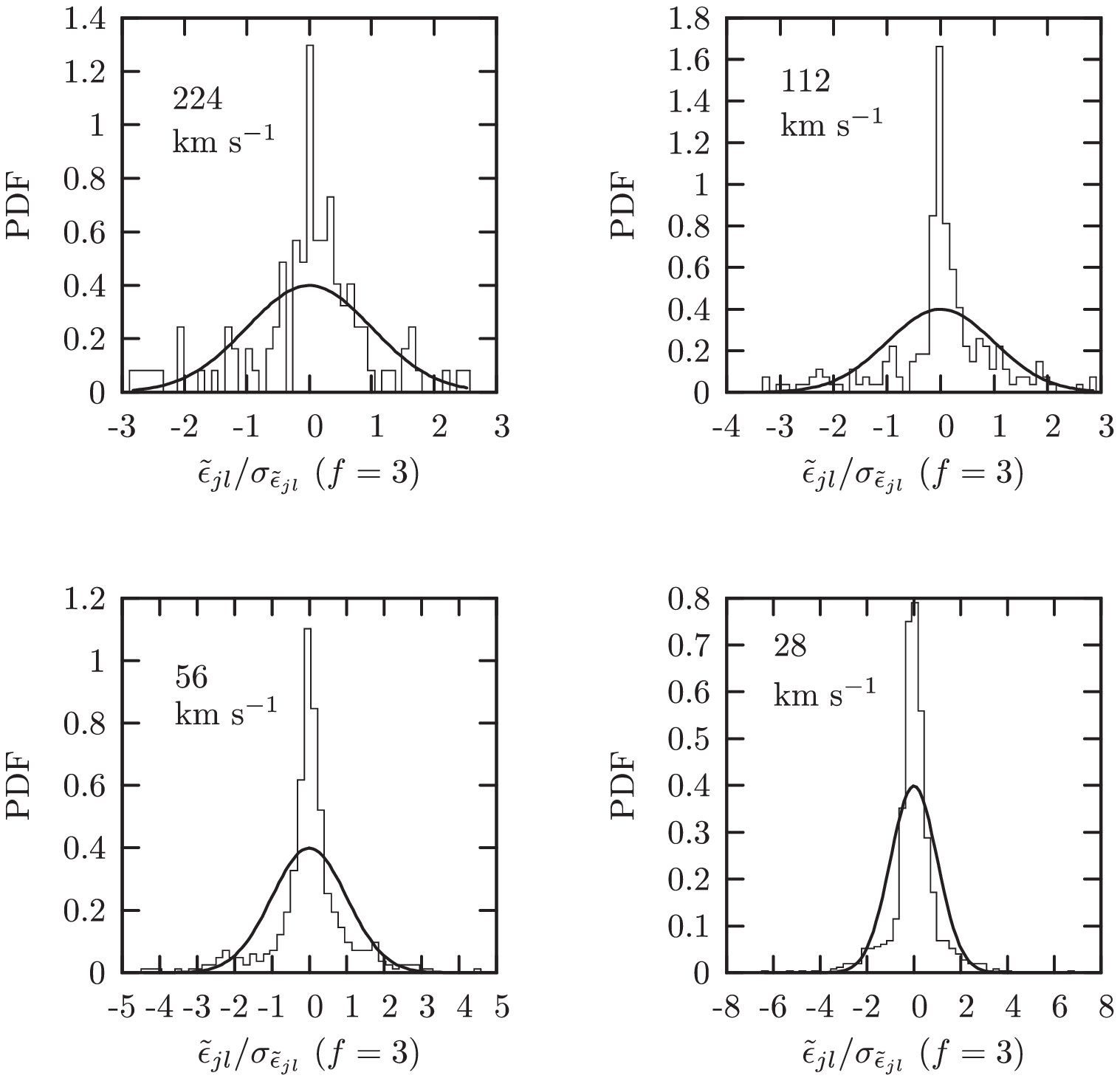} \caption{}
\end{figure}

\begin{figure}
\figurenum{3c} \epsscale{1.0} \plotone{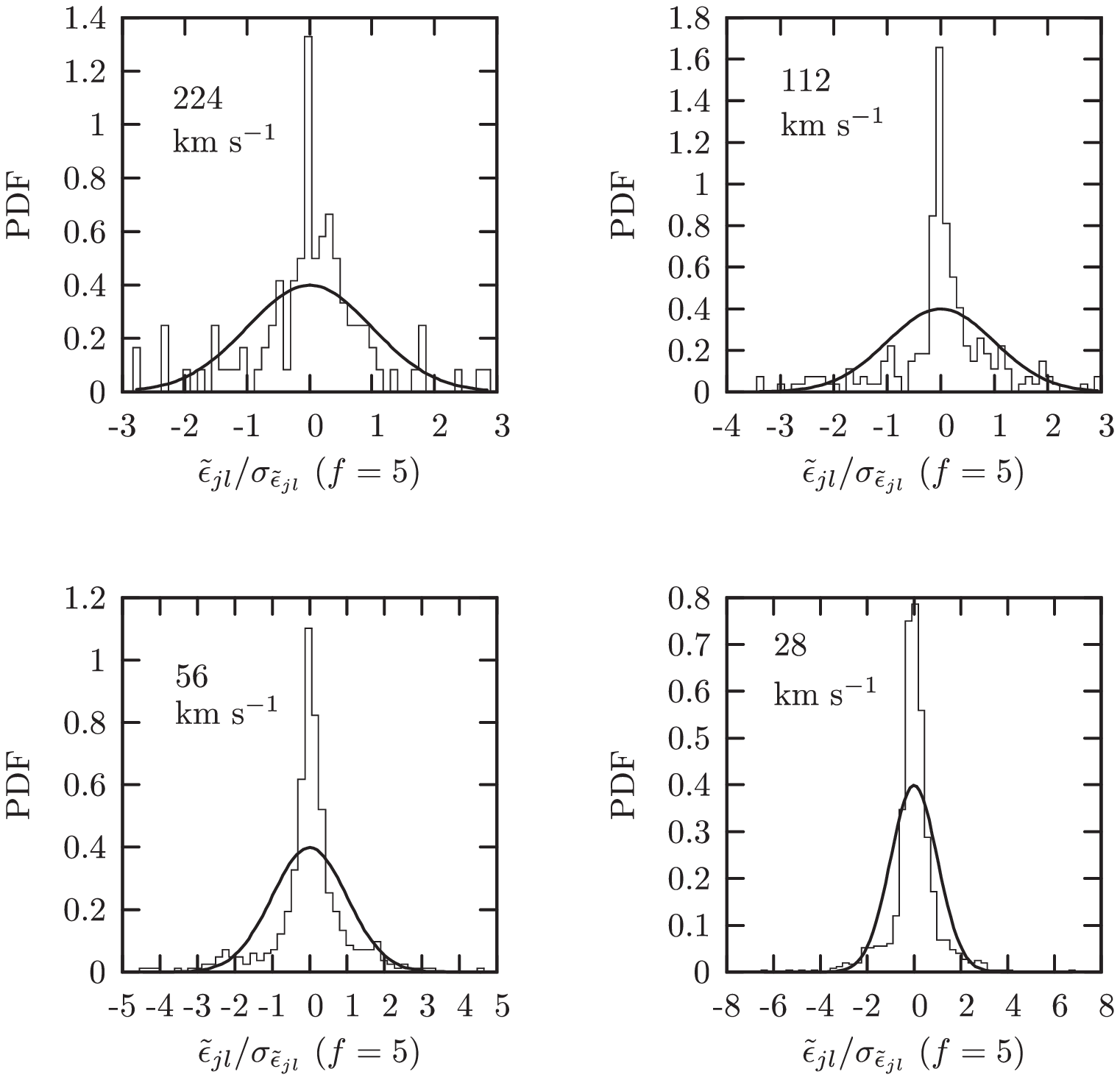} \caption{}
\end{figure}

\begin{figure}
\figurenum{4} \epsscale{1.0} \plotone{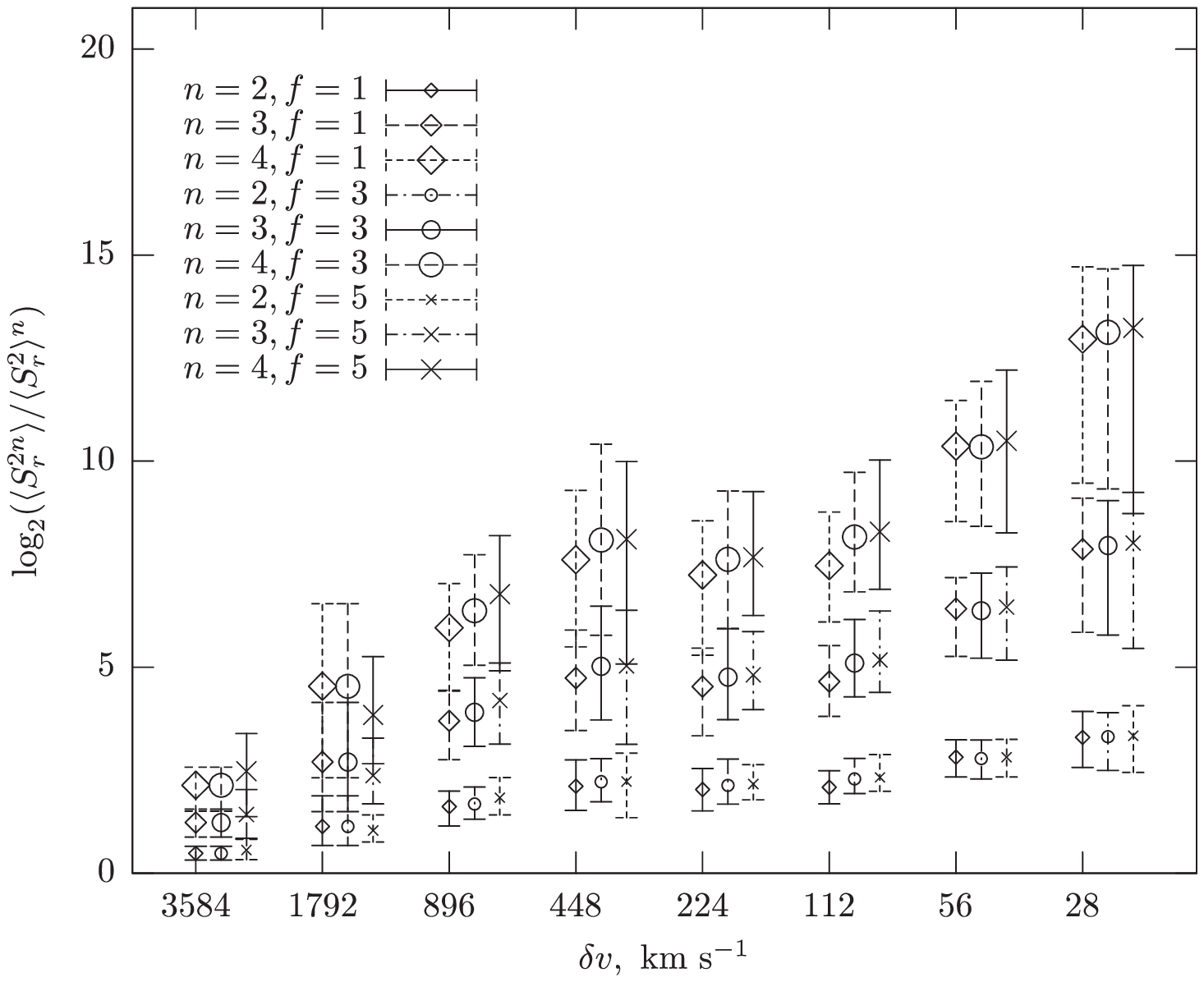} \caption{Structure
functions $\log_2[S^{2n}_j/(S^2_j)^n]$ vs. scale $\delta v$
(km~s$^{-1}$) of the data of HE2347-4342. The order $n$ is taken
to be $2, 3~{\rm and}$ 4. The parameter $f$ is taken to be $1,
3~{\rm and}$ 5. The error bars are given by bootstrap re-sampling.
For clarity, the result of $f=$ 3, and 5 are shifted slightly on
$X$-axis with respect to $f$=1.}
\end{figure}

\begin{figure}
\figurenum{5} \epsscale{1.0} \plotone{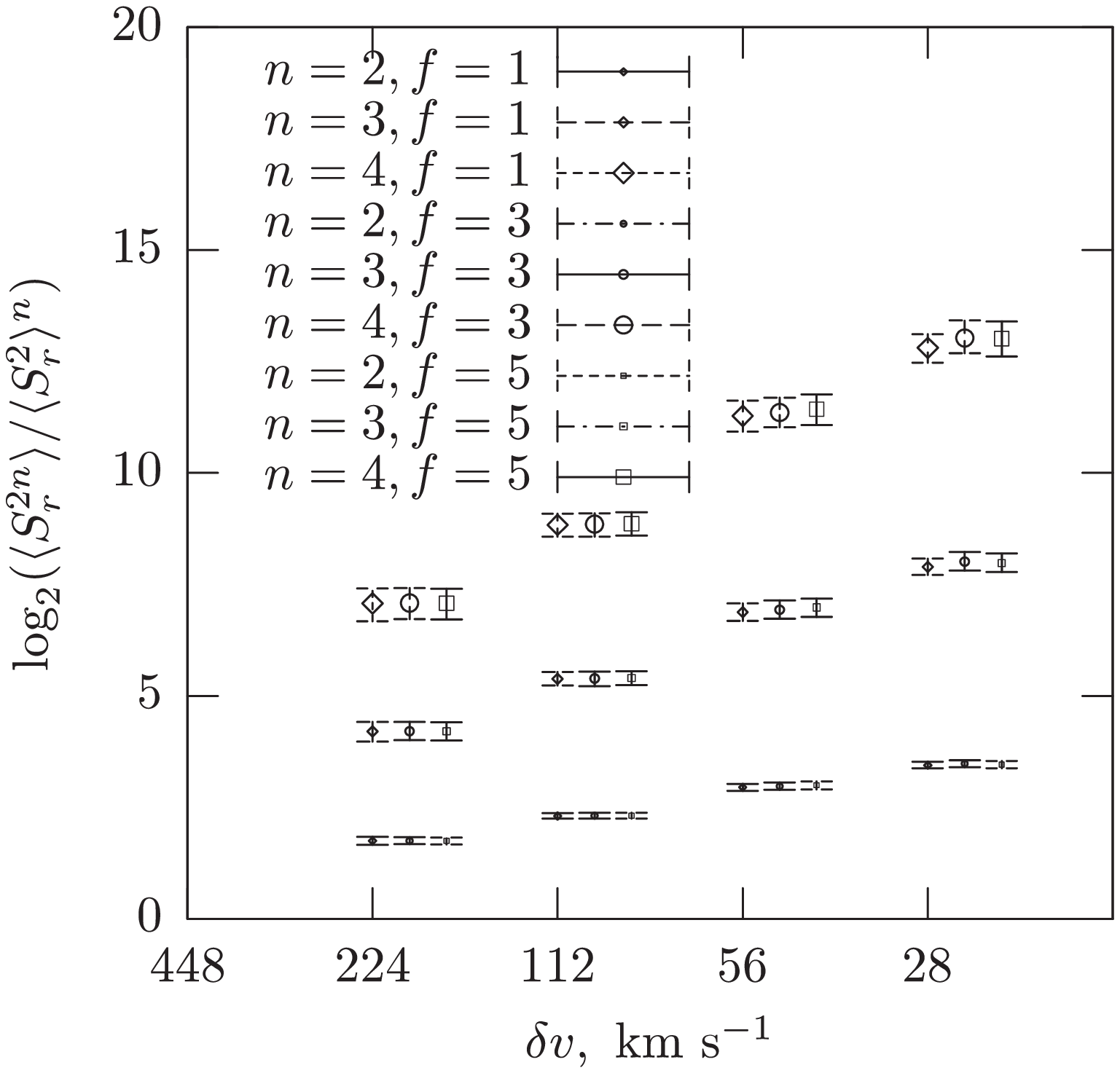} \caption{Structure
functions $\log_2[S^{2n}_j/(S^2_j)^n]$ vs. scale $\delta v$
(km~s$^{-1}$) of the mock samples of HE2347-4342. The order $n$ is
taken to be 2, 3 and 4. Parameter $f$ is taken to be 1, 3 and 5.
The error bars are the maximum and minimum of bootstrap
re-sampling. For clarity, the result of $f= 3,~{\rm and}$ 5 are
shifted slightly on $X$-axis with respect to $f=1$.}
\end{figure}

\begin{figure}
\figurenum{6} \epsscale{1.0} \plotone{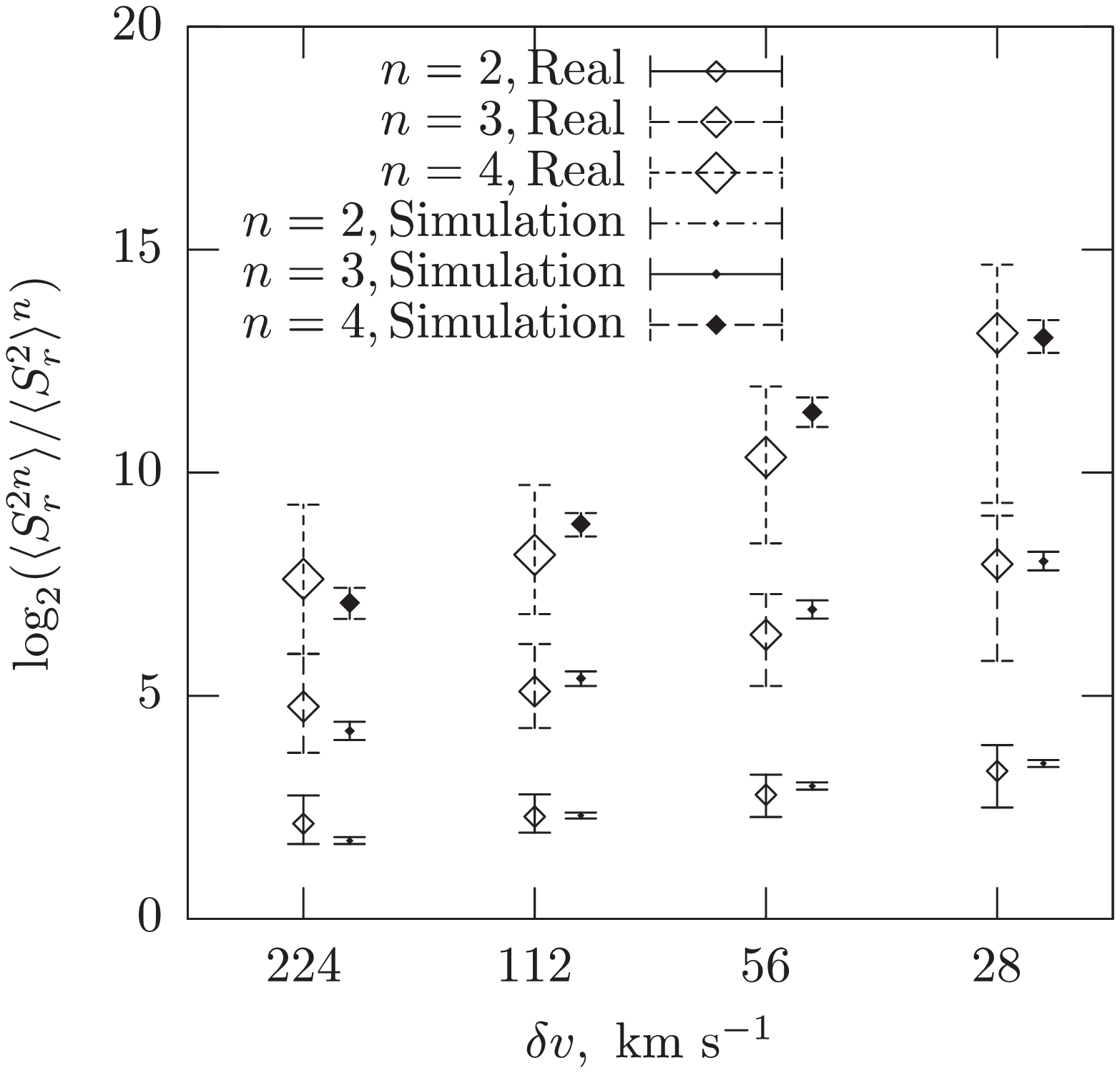} \caption{A
comparison between the structure functions of mock samples and
real data. The order $n$ is taken to be $2, 3~{\rm and~4}$. The
parameter $f$ is equal to 3. The error bars are the maximum and
minimum of bootstrap re-sampling for both types of samples.}
\end{figure}

\begin{figure}
\figurenum{7} \epsscale{1.0} \plotone{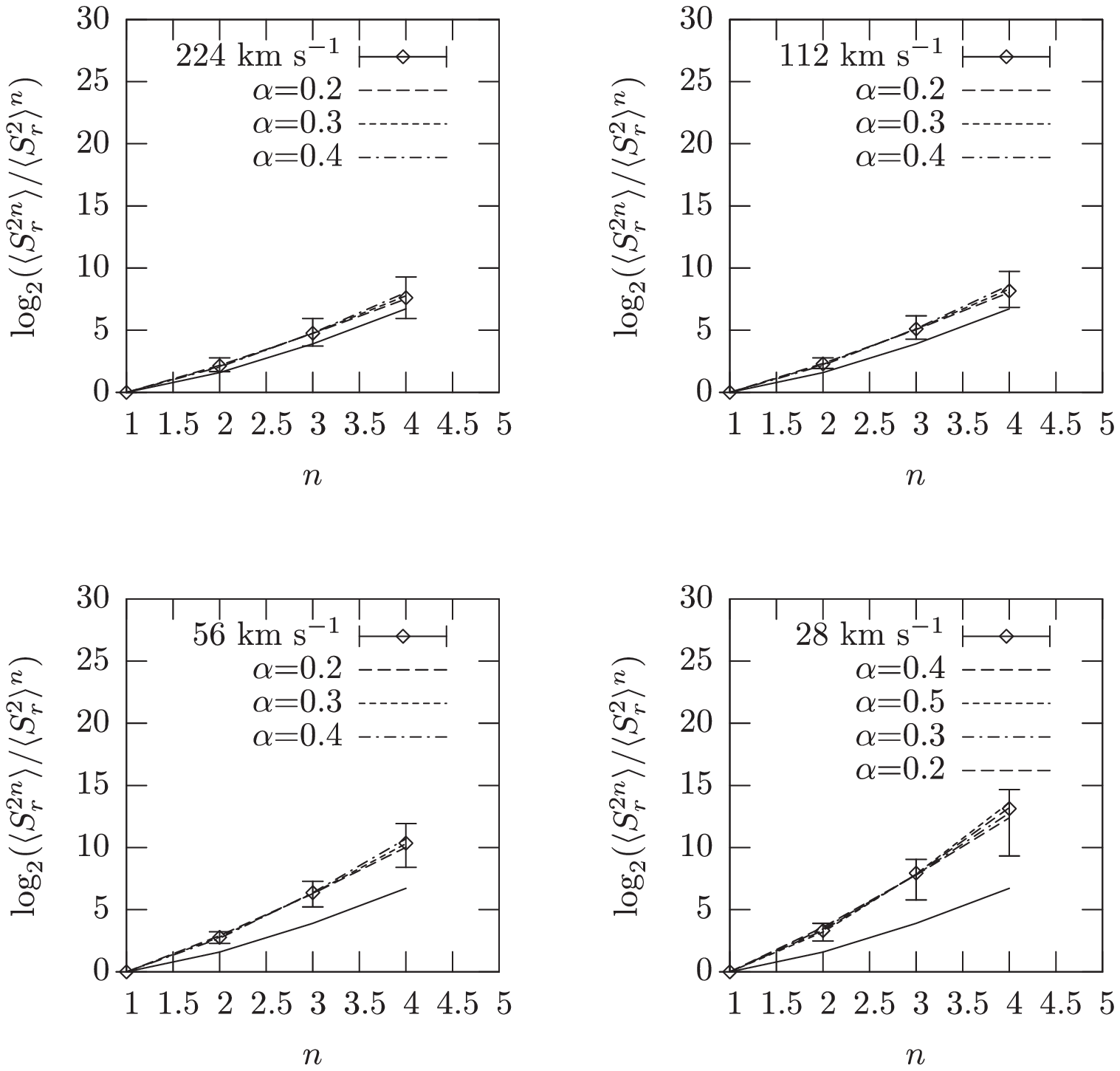} \caption{The
$\log_2[S^{2n}_j/(S^2_j)^n]$ vs. $n$ on scales of $\delta v= 224,
112, 56~{\rm and}~28$ km~s$^{-1}$ for real data of HE2347-4342.
The errors bars are given by the maximum and minimum of bootstrap
re-sampling. The fitting curves are $n^{\alpha}(n-1)$. The dotted
curves are for Gaussian field, i.e.
$\log_2[S^{2n}_j/(S^2_j)^n]=\log_2(2n-1)!!$ }
\end{figure}

\begin{figure}
\figurenum{8} \epsscale{1.0} \plotone{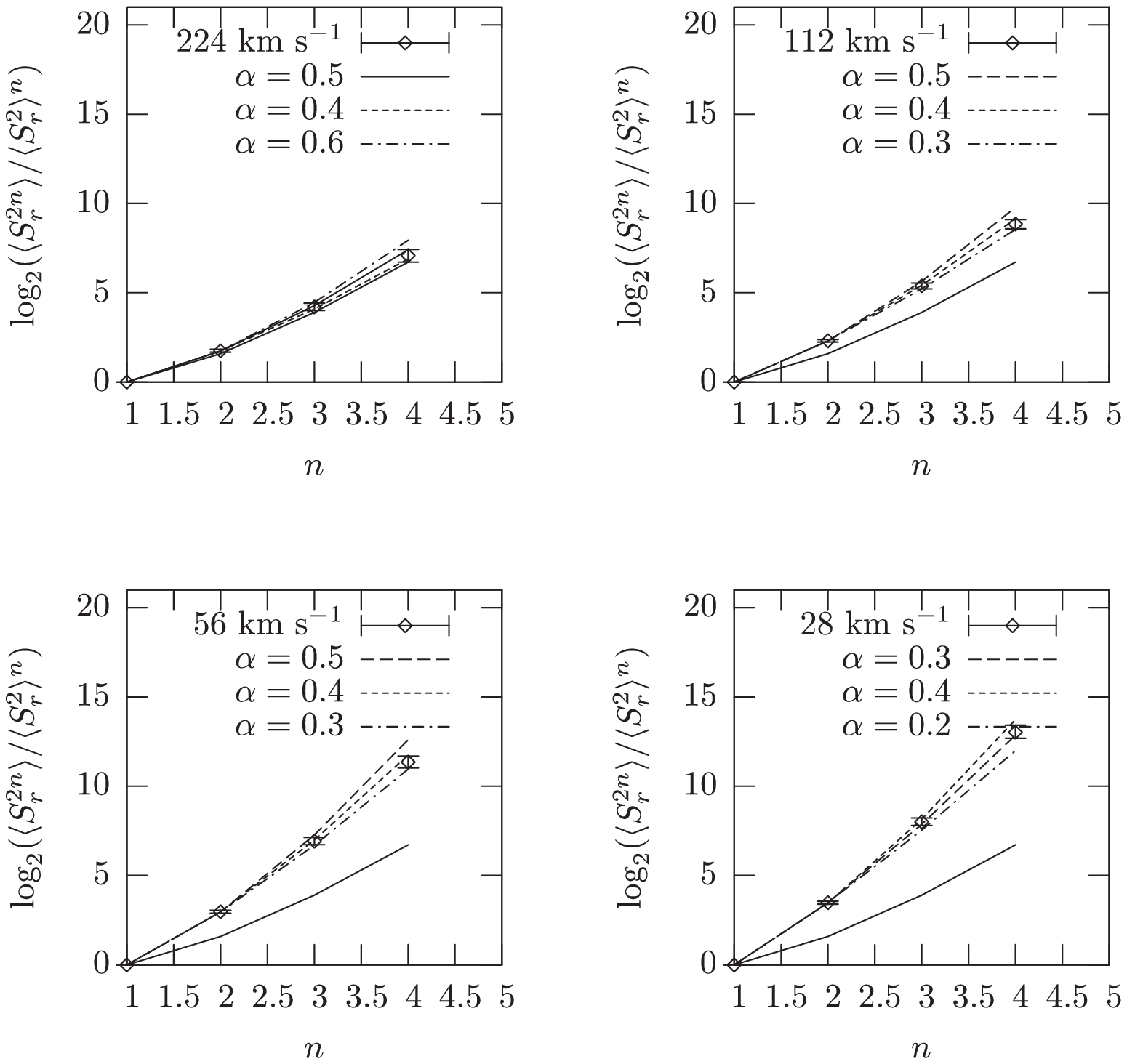} \caption{The same
as Figure 8, but for mock samples.}
\end{figure}

\end{document}